\documentclass[english,aps,manuscript]{revtex4}
\usepackage[T1]{fontenc}
\usepackage[latin9]{inputenc}
\usepackage{float}
\usepackage{textcomp}
\usepackage{amstext}
\usepackage{graphicx}
\usepackage{esint}

\makeatletter

\providecommand{\tabularnewline}{\\}

\@ifundefined{textcolor}{}
{%
 \definecolor{BLACK}{gray}{0}
 \definecolor{WHITE}{gray}{1}
 \definecolor{RED}{rgb}{1,0,0}
 \definecolor{GREEN}{rgb}{0,1,0}
 \definecolor{BLUE}{rgb}{0,0,1}
 \definecolor{CYAN}{cmyk}{1,0,0,0}
 \definecolor{MAGENTA}{cmyk}{0,1,0,0}
 \definecolor{YELLOW}{cmyk}{0,0,1,0}
}

\makeatother

\usepackage{babel}
\begin{document}

\title{Propagation of a binary signal along a chain of triangular graphane
nanoclusters}

\author{A. León}

\affiliation{Facultad de Ingeniería, Universidad Diego Portales, Santiago, Chile}
\begin{abstract}
In this paper, we study the dynamic properties of a linear array of
graphane triangular molecules that transmit a binary signal. The electronic
properties of nanoclusters are studied using calculations based on
first principles, with hybrid potentials. The dynamic of the system
is studied by solving the time-dependent Schrodinger equation. Our
results show that a linear array of these nanostructures under clock
operations, allow to transmit binary information, with a efficiency
close to unity. 
\end{abstract}
\maketitle

\section*{Introduction}

Currently, many research groups are conducting theoretical and experimental
studies, in order to find materials for the new generation of computers.
In this context, graphene is a simple bidimensional structure of carbon
atoms. In 2004 the group of Kostya Novoselov {[}1{]} succeeded in
isolating a simple layer of graphene using a technique by mechanical
exfoliation of graphite. This work represented the beginning of many
theoretical and experimental studies and their potential applications
to systems derived from graphene {[}2 \textendash{} 3{]}. A theoretical
investigation in 2007 {[}4{]} predicted a new form of graphene totally
saturated with hydrogen. The authors of this paper give the name \textquotedbl{}graphane\textquotedbl{}
to this new form derived from the graphene. The shape of this new
structure is similar to graphene, with the carbon atoms in a hexagonal
lattice and alternately hydrogenated on each side of the lattice.
In January 2009, the same group that isolated graphene in 2004 published
a paper in Science magazine reporting the hydrogenation of graphene
and the possible synthesis of graphane {[}5{]}. Since then there has
been growing scientific interest in the study of hybrid graphene-graphane
systems and their potential applications. Furthermore, has been attempted
using cellular automata to develop classical computational processes
with quantum entities. Important advances have been made with automata
based on quantum dot arrays (QCA), the idea for which was proposed
by C. Lent and collaborators {[}6{]}. The original idea was introduced
as a system of quantum corrals with four quantum dots inside it and
doped with two electrons. The electrons can tunnel through the quantum
dots, but cannot get out of the corrals that form the cells of the
automata. This architecture can propagate and process binary information
with adequate control protocols {[}7{]}. The first experimental demonstration
of the implementation of a QCA was published in 1997 {[}8{]}. A subsequent
work also demonstrated an experimental method for the implementation
of a logic gate {[}9{]}, and a shift register was also reported {[}10{]}.
These results provide good agreement between theoretical predictions
and experimental outcomes at low temperatures. Implementation at room
temperature requires working at the molecular level, and in the context
of molecular cellular automata; there are also important contributions
at the experimental level {[}11{]}. In the molecular case, the quantum
dots correspond to oxide reduction centers, and as in the case of
metallic quantum dots or semiconductors are operated with electrical
polarization. The implementation of this cellular automata architecture
is achieved with complex molecules, supported in a chemically inert
substrate. The implementation is achieved in an extremely small chain
of molecules {[}11{]}. Another implementation at room temperature
corresponds to an array of magnetic quantum dots that can propagate
magnetic excitations to process digital information {[}12{]}. These
systems use the magnetic dipolar interaction among particles whose
size is at the submicrometer scale. A theoretical study was published
about the behavior of cellular automata composed of an array of polycyclic
aromatic molecules {[}13{]}, using the polarization of electronic
spin. In this work it is established that by increasing molecules
in one of the directions of the plane, forming graphene nanoribbons,
binary information can be transmitted at room temperature. In another
work, we investigate the properties of a molecular array of graphane
{[}14{]}. This study established that graphane nanoclusters, with
two quantum dots, are eligible to transmit binary information. In
this paper we investigate the dynamic behavior of a linear array of
triangular graphane molecules, under clock operations to propagate
binary information.

\section*{Triangular Graphane Molecules}

We can consider a cell in the automaton that contains three quantum
dots to provide an additional state. In this way, each cell would
have the polarization state P = +1, if the hole (or electron) is found
in one of the two active quantum dots, P = -1, if it is found in the
other active dot and P = 0 if it is found in the null point {[}12{]}.
This idea of a clocked cell has been studied with the theoretical
model of the Aviram molecule {[}17{]}. This model poses the possibility
of activating and deactivating each cell by using an electrical field.
It is important to consider this type of cell in our structures, given
that an appropriate line of clocked cells forms a shift register,
a key component in QCA circuitry. A continuously varying clocking
wave can move bits of information smoothly along the line. Power gain
is essential for practical operations of systems that process information.
A result of the dissipative effects, a part of the signal sent from
one sector to another of a circuit is lost. In conventional electronics,
the energy of the signal is reinforced with energy provided by the
power source. In QCA technology devices, this energy is supplied by
the clock. In this way, if the energy supplied to a particular cell
falls below the operational threshold, the clock supplies additional
energy so that the signal flows with a gain greater than one {[}18{]}.
It has been theoretically and experimentally verified that this effect
is automatic as a consequence of the interactions among the systems.
Power gains greater than 3 have been measured {[}19{]}. In this investigation
we will study the properties and dynamic behavior of cells with three
quantum dots to estimate power gain.

The dynamic properties of molecular array depend on the electronic
properties of individual cells. The molecules studied in this work
are triangular molecules of graphane (Figure 1), similar to triangular
nanoclusters of graphene, saturated with hydrogen atoms, except at
the corners (quantum dots). In those corners there are two unpaired
electrons in each quantum dot and in the case of the molecular double
cation (Figure 1b), there is an unpaired electron that can be located
in a quantum dot or in the others. Our results obtained using methods
based on First-principles calculations, shows that molecular double
cation has a state of minimum energy with the electron located in
one of the quantum dots (or superposition states in small molecules)
and it has an excited state when the electron is positioned in any
of the other quantum dots. In all our calculations the structure is
relaxed by means of the method \textquotedblleft{}Quasi-Newton Approach\textquotedblright{}
{[}15{]} and we adopt Slater type orbitals and triple zeta polarized
basis (TZP) wiht hybrids method B3LYP\textasteriskcentered{}, by using
the ADF software {[}16{]}. In the previous work {[}14{]}, we has established
the best hybrid method, (First-principles calculation), for this type
of the molecules, this method is B3LYP{*}.

\begin{figure}[H]
\includegraphics[scale=0.35]{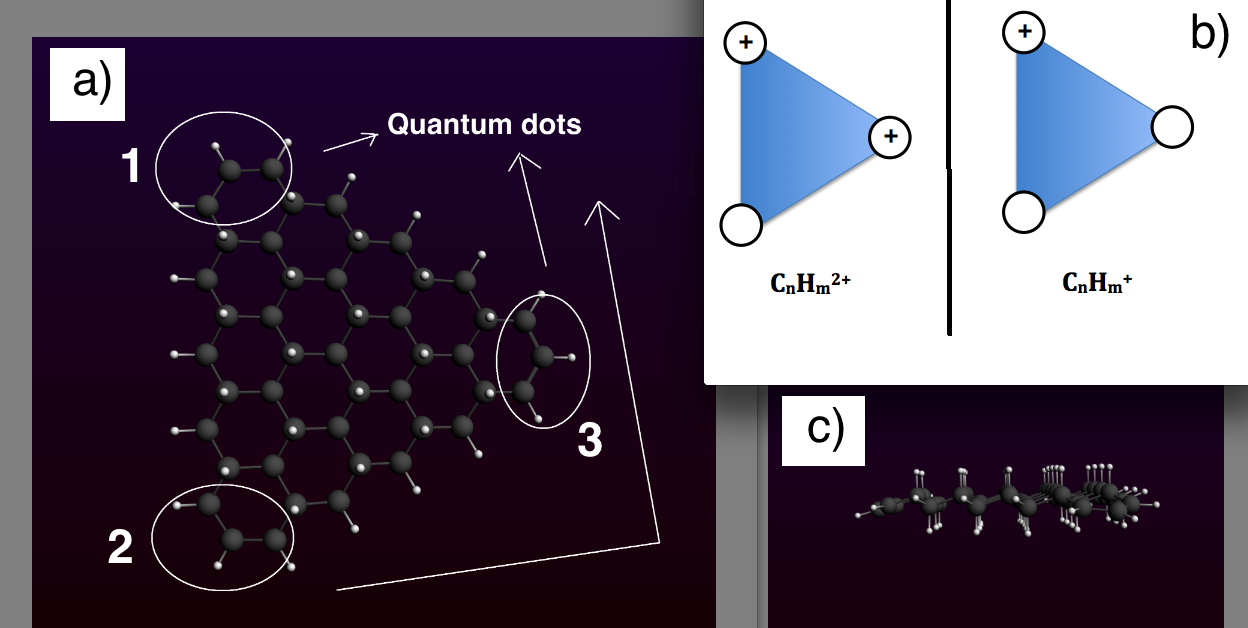}

Figure 1: a) Triangular graphane molecule with three quantum dots.
b) Scheme of the molecular double cation and molecular single cation.
c) Scheme of the triangular graphane molecule in the plane.

\end{figure}

In the molecular double cation, we define one active state and other
inactive state. In the active state, we can define two logic states
(0 and 1) and we can activate the molecule, with a electric field
(clock field). Figure 2, shows the two logic states and inactive state.
When the clock field is applied in the direction $-x$, the molecule
has a state of minimum energy with the electron located in the quantum
dot 3 (inactive state). The Figure 2c show the HOMO in the inactive
state. When the clock field is applied in the direction $x$ the molecule
has a state of minimum energy with the electron located in the quantum
dot 1, or located in the quantum dot 2. Figures 2a and 2b shows this
situation.

\begin{figure}[H]
\includegraphics[scale=0.3]{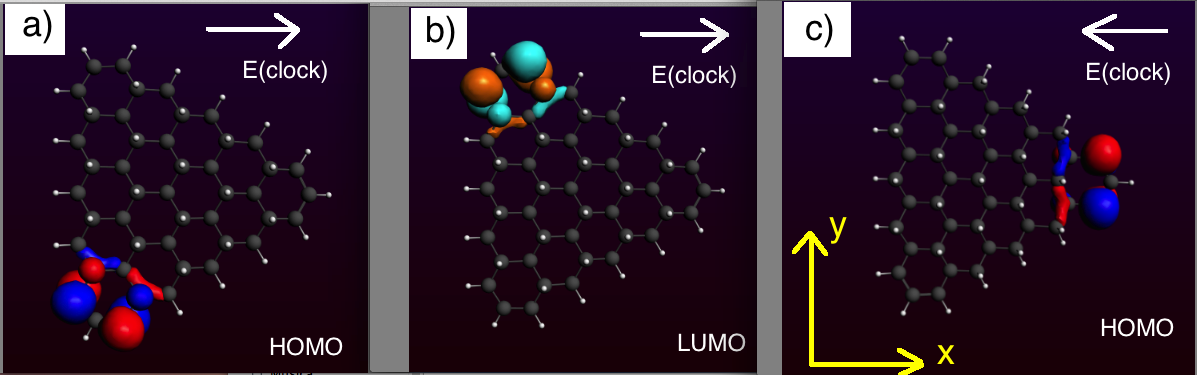}

Figure 2. a) Logic state 0, b) logic state 1 and c) inactive state.
\end{figure}

A very important parameter to determine the dynamic properties of
molecular array, is the energy of tunneling between quantum dots of
the molecule. For molecular double cation, in the active state, we
can define this energy as $\gamma=\frac{E(LUMO)-E(HOMO)}{2}$. The
table 1 show the $\gamma$ values, for some triangular molecules.
The $L$ parameter correspond to the side of the equilateral triangule. 

\begin{table}[H]
\begin{tabular}{|c|c|c|}
\hline 
$L\,\left(\textrm{\AA}\right)$ & $E\left[Clock\right]\,\left(\frac{Hartree}{e\: Bohr}\right)$ & $\gamma\,\left(mHartree\right)$\tabularnewline
\hline 
\hline 
7.42 & 0.008 & 9.10\tabularnewline
\hline 
7.42 & 0.010 & 9.25\tabularnewline
\hline 
7.42 & 0.020 & 9.40\tabularnewline
\hline 
7.42 & 0.040 & 9.69\tabularnewline
\hline 
12.58 & 0.010 & 3.80\tabularnewline
\hline 
12.58 & 0.013 & 4.00\tabularnewline
\hline 
\end{tabular}

\caption{Energy of tunneling between quantum dots of the molecules, in the
active state.}

\end{table}

Our results shows that the $\gamma$ parameter depend of the magnitude
of the clock field and of the molecular size. This result imply that
we can dispose of a fine control on the dynamics property of the molecular
array.

\section*{Propagation of binary signal}

In this section we analyse the dynamic response of the molecular array
when binary information is transmitted from one extreme to the other
of the system. Let us consider a system consisting of $N$ cells,
as shown Figure 3. The first cell acts as a driver and its polarization
is externally manipulated. For the remaining $N-1$ active cells,
we solve the time-dependent Schrödinger equation using the occupation
number Hamiltonian. Most of the work which exists in the literature
on multiple quantum dot systems is based on occupation number (Hubbard-like)
formalism {[}20\textendash{}22{]}. This type of Hamiltonian is rather
straightforward to implement and requires very limited computational
resources. Each quantum dot and its interaction with the other dots
are described by means of a few phenomenological parameters, such
as the tunneling energy, the dot confinement energy, the onsite electrostatic
interaction. Such descriptions are successful in capturing the overall
behavior of the system, and in providing a qualitative understanding
of the underlying physics. For the i-cell (in the active state) in
the automaton, the Hamiltonian can be written: 

\begin{equation}
H=V_{1}\hat{n}_{1}+V_{2}\hat{n}_{2}-\gamma\left(\hat{a}_{1}^{\dagger}\hat{a}_{2}+\hat{a}_{2}^{\dagger}\hat{a}_{1}\right)
\end{equation}

In this equation the subscripts 1 and 2 denote the upper and lower
quantum dot respectively (Fig. 1a). The terms $\hat{n},\;\hat{a}^{\dagger}$
and $\hat{a}$ correspond to the number, creation and annihilation
operators, respectively. The parameter $\gamma$ corresponds to the
tunneling energy between the quantum dots of the molecule and it was
defined in the previous section. The terms $V_{1}$ and $V_{2}$ represent
the Coulomb interaction between the charge in the dot 1 or 2 and the
other cells. These terms take into account the interaction with neighboring.
The vector representing the state of the i-cell is written as: 

\begin{equation}
\left|\psi\left(t\right)\right\rangle =c_{1}\left(t\right)\left|1\right\rangle +c_{2}\left(t\right)\left|2\right\rangle 
\end{equation}

where $\left|1\right\rangle $ and $\left|2\right\rangle $ are the
charge wave functions as it is in the upper quantum dot and in the
lower, respectively. Using the Hamiltonian given by Eq. (1), the dynamics
is obtained from the equation:

\begin{equation}
i\frac{\partial\left|\psi\left(t\right)\right\rangle }{\partial t}=H\left|\psi\left(t\right)\right\rangle 
\end{equation}

By replacing the wave vector (2) in the Schrödinger equation (3),
a system of differential equations for the coefficients are obtained
which we solve numerically with given initial conditions. In our study,
the value of polarization P = +1 (in atomic units) means the hole
(positive charge) is in the upper quantum dot and P = \textminus{}1,
means it is on the lower quantum dot. 

\begin{figure}[H]
\includegraphics[scale=0.5]{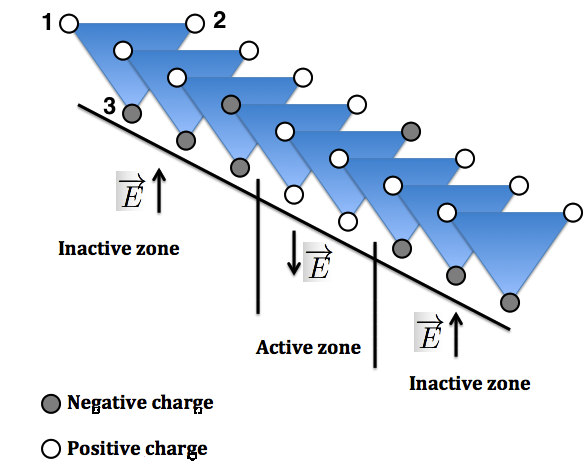} 

Figure 3: Scheme of the molecular array with the actives zones and
inactives zones.

\end{figure}

\section*{Results and conclusions}

Clock operation is defined as follows: In a time period $\Omega$,
only two cells are in the active state, the rest of the system is
in the idle state. Each cell remains in the active state for a time
equal to $2\Omega$. When a period ends and the next begins, the electric
field causes it to disable one of the cells and activate the next.
This situation is depicted in Figure 4. We define the time parameter
as: $\Omega=\frac{8\pi}{\triangle\:\gamma}$, where $\triangle$ corresponds
to the step of the algorithm used to solve the system of equations
and $\gamma$ parameter was defined above. The study of the properties
of the molecule, shows us that we can modulate the $\gamma$ parameter
with the electric field (clock). During clock operation, we change
the magnitude of the electric field on a time interval of $\left(t_{clock}\right)$,
to modify the $\gamma$ parameter.

\begin{figure}[H]
\includegraphics[scale=0.35]{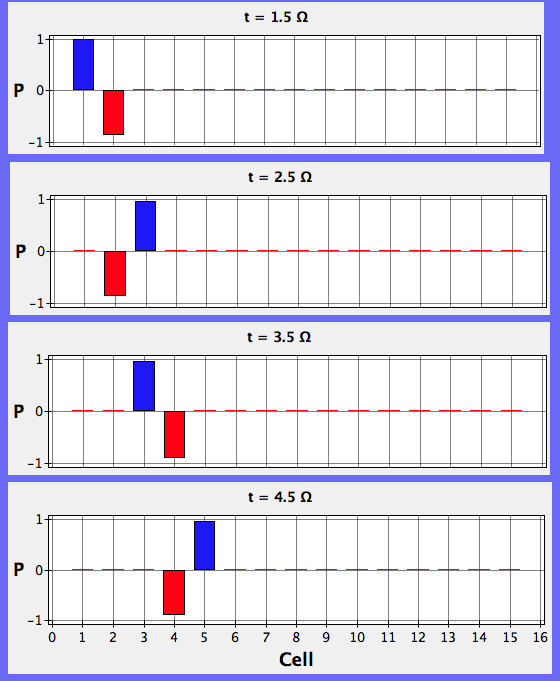}

Figure 4. Scheme clock operation of molecular array with 15 cells.
In the $x$ axis, the cells that form the array are presented. In
the $y$ axis, the polarization of each of the cells (in atomic units)
is represented.
\end{figure}

The first system studied, is formed by 15 triangular molecules of
size 1.26 nm (Figure 1a). The separation distance between the molecules
is 1.00 nm. Figure 4 shows results, for $t_{clock}=0.9\Omega$, $\Omega=3.04\; ps$,
$\gamma_{max}=0.020\; Hartree$ and $\gamma_{min}=0.001\; Hartree$.
Figure 4a shows $\gamma$ parameter for each cell (The colors in the
figure correspond to the $\gamma$ parameter of the cells). Figure
4b shows polarization of each cell over time and Figure 4c shows the
polarization of molecular array for some moment. We see that the digital
signal arrives slightly degraded at the other end of the lineal molecular
array. We define $\eta=\frac{P_{out}}{P_{in}}$ as a measure of efficient,
where $P_{in}$ is the polarization of the cell 1 and $P_{out}$ is
the polarization of last cell. By definition, the polarization of
first cell is equal to one. We calculate the polarization of last
cell using:

\begin{equation}
P=\frac{1}{\Omega}\intop_{N\Omega}^{\left(N+1\right)\Omega}P\left(t\right)dt
\end{equation}
where $P\left(t\right)$ is polarization over time of last cell and
$N$ is the cells number. 

\begin{figure}[H]
\includegraphics[scale=0.35]{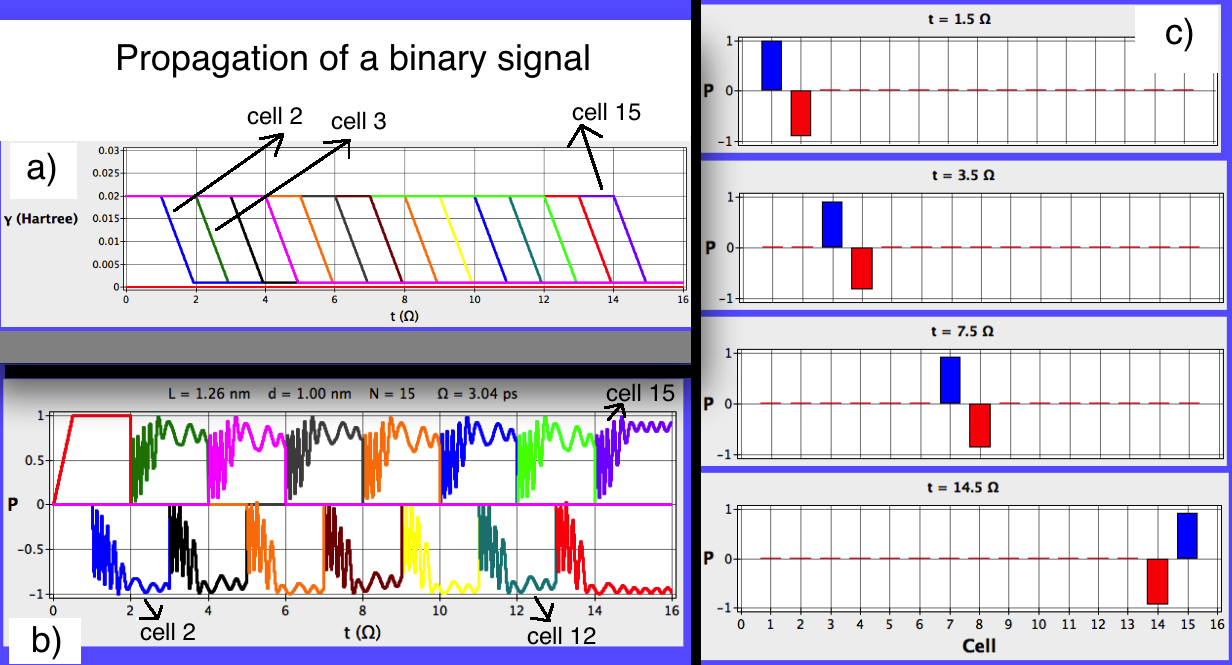}

Figure 5: a) Scheme of $\gamma$ parameter's change over time. b)
Polarization of the molecular array over time. c) Propagation of a
bit information.

\end{figure}

In this first simulation the value of $\eta$ is 0.874, close to the
unit value. Our study verified that the efficiency $\eta$ increases
if $t_{clock}$ decreases, for this physical configuration of the
molecules. Figure 6 shows the results for the same system of Figure
5, but with $t_{clock}=0.09\Omega$. We can observe that the polarization
of each cell is higher (in magnitude) than in the previous case and
the value of the efficiency in this case is 0.933.

\begin{figure}[H]
\includegraphics[scale=0.35]{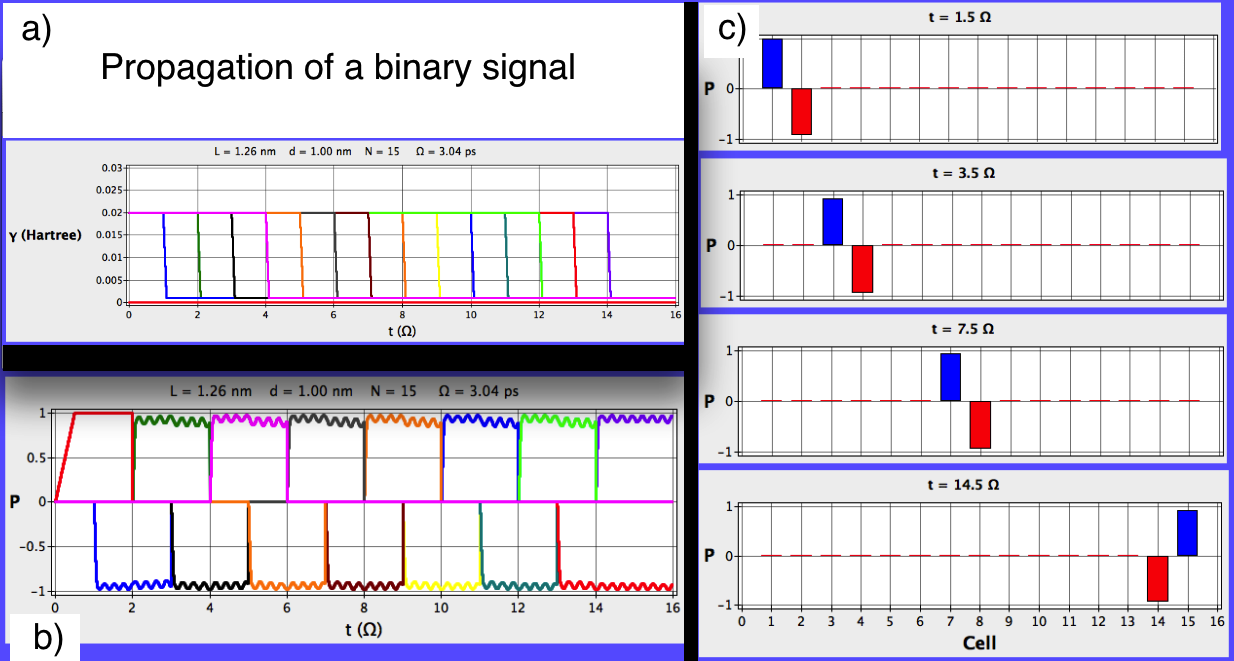}

Figure 6: a) a) Scheme of $\gamma$ parameter's change over time.
b) Polarization of the molecular array over time. c) Propagation of
a bit information.

\end{figure}

We perform simulations for different molecule sizes, separation distances,
tunneling energy and duration of the operation clock. We were able
to verify that the efficiency depends on all these parameters and
for chains whose length is greater than 15 cells, the value of the
efficiency is close to unity (greater than 90\%). We conclude that
a linear array of triangular graphane molecules, with three quantum
dots, digital information can be transmitted long distances without
damaging the bit of information, as long as we make a proper clock
operation.

In this work we developed a proof of concept to show that a linear
arrangement of graphane nanoclusters efficiently transmits one bit
of information an appreciable distance. From this theoretical study,
one could design an experimental protocol to demonstrate the behavior
of graphane in such applications.

\section*{Acknowledgment}

The author acknowledge the financial support of FONDECYT program grant
11100045.

\section*{References}

{[}1{]} K. S. Nonoselov, A. K. Geim, S. V. Morozov, D. Jiang, Y. Zhang,
S. V. Dubonos, I. V. Grigorieva, A. A. Firsov, Science 306, 666 (2004). 

{[}2{]} A. K. Geim, Sience 324, 1530 (2009). 

{[}3{]} A. H. Castro Neto, F. Guinea, N. M.R. Peres, K. S. Novoselov,
A. K. Geim, Rev. Mod. Phys. 81, 109 (2009). 

{[}4{]} J. O. Sofo, A. S. Chaudhari, G. D. Barber, Phys. Rev. B 75,
153401 (2007). 

{[}5{]} D. C. Elias et al, Science 323, 610 (2009).

{[}6{]} C. S. Lent, P. D. Tougaw, W. Porod, and G. H. Bernstein, Nanotechnology
4, 49 (1993). 

{[}7{]} A. I. Csurgay, W. Porod, and C. S. Lent, IEEE Trans. On Circuits
and Systems I 47, 1212 (2000). 

{[}8{]} A. O. Orlov, I. Amlani, G. H. Bernstein, C. S. Lent, and G.
L. Snider, Science 277, 928 (1997). 

{[}9{]} I. Amlani, A. O. Orlov, G. Toth, G. H. Bernstein, C. S. Lent,
and G. L. Snider, Science 284, 289 (1999). 

{[}10{]} R. K. Kummamuru, A. O. Orlov, R. Ramasubramaniam, C. S. Lent,
G. H. Bernstein, and G. L. Snider, IEEE Trans. On Electron Devices
50, 1906 (2003). 

{[}11{]} J. Jiao, G. J. Long, F. Grandjean, A. M. Beatty, T. P. Fehlner,
J. Am. Chem. Soc. 125, 7522 (2003). 

{[}12{]} M. Macucci, Quantum Cellular Automata (Imperial College Press,
London, 2006). 

{[}13{]} A. León, Z. Barticevic and M. Pacheco, Appl. Phys. Lett.
94, 173111 (2009).

{[}14{]} A. León and M. Pacheco, Phys. Lett. A 375 (2011) 4190-4197.

{[}15{]} L. Versluis, The Determination of Molecular Structures by
the HFS Method, Uni- versity of Calgary, 1989.

{[}16{]} http://www.scm.com/. 

{[}17{]} C. S. Lent and P. D. Tougaw, Proc. Of the IEEE 85, 541 (1997). 

{[}18{]} J. Timler and C. S. Lent, J. Appl. Phys. 91, 823 (2002). 

{[}19{]} A. O. Orlov et al, Appl. Phys. Lett. 81, 1332 (2002).

{[}20{]} P.D. Tougaw, C.S. Lent, W. Porod, J. Appl. Phys. 74 (1993)
3558. 

{[}21{]} P.D. Tougaw, C.S. Lent, J. Appl. Phys. 75 (1994) 1818. 

{[}22{]} C.A. Stafford, S. Das Sarma, Phys. Rev. Lett. 72 (1994) 3590.
\end{document}